\documentclass[aps,prl,twocolumn,superscriptaddress,floatfix]{revtex4}%
\usepackage{graphicx}
\usepackage{dcolumn}
\usepackage{bm}
\usepackage{hyperref}
\usepackage{amssymb}
\usepackage{amsmath}
\usepackage{amsfonts}
\usepackage{amssymb}
\usepackage{color}%
\begin{document}
\title{Observation of interference between resonant and detuned STIRAP in the adiabatic creation of $^{23}$Na$^{40}$K molecules}
\author{Lan Liu*}
\affiliation{Hefei National Laboratory for Physical Sciences at Microscale and Department
of Modern Physics, University of Science and Technology of China, Hefei,
Anhui 230026, China}
\affiliation{Shanghai Branch, CAS Center for Excellence and Synergetic Innovation Center in Quantum
Information and Quantum Physics, University of Science and Technology of
China, Shanghai 201315, China}
\author{De-Chao Zhang*}
\affiliation{Hefei National Laboratory for Physical Sciences at Microscale and Department
of Modern Physics, University of Science and Technology of China, Hefei,
Anhui 230026, China}
\affiliation{Shanghai Branch, CAS Center for Excellence and Synergetic Innovation Center in Quantum
Information and Quantum Physics, University of Science and Technology of
China, Shanghai 201315, China}
\author{Huan Yang}
\affiliation{Hefei National Laboratory for Physical Sciences at Microscale and Department
of Modern Physics, University of Science and Technology of China, Hefei,
Anhui 230026, China}
\affiliation{Shanghai Branch, CAS Center for Excellence and Synergetic Innovation Center in Quantum
Information and Quantum Physics, University of Science and Technology of
China, Shanghai 201315, China}
\author{Ya-Xiong Liu}
\affiliation{Hefei National Laboratory for Physical Sciences at Microscale and Department
of Modern Physics, University of Science and Technology of China, Hefei,
Anhui 230026, China}
\affiliation{Shanghai Branch, CAS Center for Excellence and Synergetic Innovation Center in Quantum
Information and Quantum Physics, University of Science and Technology of
China, Shanghai 201315, China}
\author{Jue Nan}
\affiliation{Hefei National Laboratory for Physical Sciences at Microscale and Department
of Modern Physics, University of Science and Technology of China, Hefei,
Anhui 230026, China}
\affiliation{Shanghai Branch, CAS Center for Excellence and Synergetic Innovation Center in Quantum
Information and Quantum Physics, University of Science and Technology of
China, Shanghai 201315, China}
\author{Jun Rui}
\affiliation{Hefei National Laboratory for Physical Sciences at Microscale and Department
of Modern Physics, University of Science and Technology of China, Hefei,
Anhui 230026, China}
\affiliation{Shanghai Branch, CAS Center for Excellence and Synergetic Innovation Center in Quantum
Information and Quantum Physics, University of Science and Technology of
China, Shanghai 201315, China}
\author{Bo Zhao}
\affiliation{Hefei National Laboratory for Physical Sciences at Microscale and Department
of Modern Physics, University of Science and Technology of China, Hefei,
Anhui 230026, China}
\affiliation{Shanghai Branch, CAS Center for Excellence and Synergetic Innovation Center in Quantum
Information and Quantum Physics, University of Science and Technology of
China, Shanghai 201315, China}
\author{Jian-Wei Pan}
\affiliation{Hefei National Laboratory for Physical Sciences at Microscale and Department
of Modern Physics, University of Science and Technology of China, Hefei,
Anhui 230026, China}
\affiliation{Shanghai Branch, CAS Center for Excellence and Synergetic Innovation Center in Quantum
Information and Quantum Physics, University of Science and Technology of
China, Shanghai 201315, China}

\begin{abstract}{Stimulated Raman adiabatic passage (STIRAP) allows to efficiently transferring the populations between two discrete quantum states and has been used to prepare molecules in their rovibrational ground state. In realistic molecules, a well-resolved intermediate state is usually selected to implement the resonant STIRAP. Due to the complex molecular level structures, the detuned STIRAP always coexists with the resonant STIRAP and may cause unexpected interference phenomenon. However, it is generally accepted that the detuned STIRAP can be neglected if compared with the resonant STIRAP. Here we report on the first observation of interference between the resonant and detuned STIRAP in the adiabatic creation of $^{23}$Na$^{40}$K ground-state molecules. The interference is identified by observing that the number of Feshbach molecules after a round-trip STIRAP oscillates as a function of the hold time, with a visibility of about 90\%.  This occurs even if the intermediate excited states are well resolved, and the single-photon detuning of the detuned STIRAP is about one order of magnitude larger than the linewidth of the excited state and the Rabi frequencies of the STIRAP lasers. Moreover, the observed interference indicates that if more than one hyperfine level of the ground state is populated, the STIRAP prepares a coherent superposition state among them, but not an incoherent mixed state. Further, the purity of the hyperfine levels of the created ground state can be quantitatively determined by the visibility of the oscillation.}
\end{abstract}
\maketitle
\renewcommand{\thefootnote}{\fnsymbol{footnote}} \footnotetext[1]{%
These authors contributed equally to this work.} \renewcommand{\thefootnote}{%
\arabic{footnote}}

Ultracold polar molecules offer great opportunities to study ultracold
chemistry \cite{Ospelkaus2010a,Knoop2010,Rui2017,Yang2019}, quantum simulation \cite{Micheli2006,Baranov2012}, quantum computation \cite{Demille2002,Rabl2006} and to perform precision measurements \cite{Demille2008,Zelevinsky2008,Baron2014}. A precise control of the electronic,
vibrational, rotational and hyperfine states of polar molecules is of great importance to these applications \cite{Carr2009,Quemener2012}. Stimulated Raman adiabatic passage (STIRAP) allows for a coherent population transfer between two molecular internal quantum states \cite{Bergmann1998,Vitanov2017}, and has been employed to create and detect the ultracold alkali-metal-diatomic molecules in the rovibrational ground state \cite{Ni2008,Molony2014,Takekoshi2014,Park2015,Guo2016,Rvachov2017,seesselberg2018,Ospelkaus2010a,Yang2019}. The principle of STIRAP can be understood using a three-level model, which involves the Feshbach state, the intermediate electronic excited state, and the molecular rovibrational ground state \cite{Ni2008, Bergmann1998}.
However, in realistic molecules, both the electronic excited state and
the rovibrational ground state have many hyperfine levels \cite{Ospelkaus2010b,Park2015b,Aldegunde2017}.
These complex level structures raise several open questions on  how STIRAP works in a realistic molecule system.

The first open question is about the intermediate electronic excited state. In a molecule, each rovibrational level of the electronic excited state contains many hyperfine levels. Therefore, in practice, a well-resolved hyperfine level of the excited
state is selected as the intermediate state by setting the single-photon detuning to be
zero \cite{Ni2008,Molony2014,Takekoshi2014,Park2015,Guo2016,Rvachov2017}. This hyperfine excited state should be dominantly coupled to one of the
hyperfine levels of the ground state, and thus a resonantly coupled three-level system is selected.  It is usually assumed that if the energy difference between the selected and the adjacent
hyperfine level of the excited state is much larger than the natural linewidth of the excited
states and the Rabi frequencies of the STIRAP lasers, then only the resonant STIRAP plays a dominant role, and the other excited states can be
safely neglected.

STIRAP can also proceed via other hyperfine levels of the excited state with a single-photon
detuning \cite{Vitanov1999,Bergmann1998,Vitanov2017,seesselberg2018}. In previous works, detuned STIRAPs are usually neglected, once a well-resolved
hyperfine excited state has been selected for the resonant STIRAP. However, detuned STIRAP always exists in the realistic molecule systems and may contribute the unexpected transfer channels. Since STIRAP is a coherent process,
the different channels due to the resonant and detuned STIRAP may interfere with each other and cause
unexpected interference phenomenon.
Despite that STIRAPs in multi-level systems have been intensively studied in previous works~\cite{Bergmann1998,Vitanov2017} and the various scenarios, such as the resonant and off-resonant multistate chains~\cite{Shore1991,Pillet1993,Vitanov1998}, the multiple intermediate states \cite{Vitanov1999}, and the multiple ground states~\cite{Martin1995, Martin1996}, have been investigated, the interference between resonant STIRAP and detuned STIRAP has not been observed before.


The second open question is about the hyperfine levels of the ground state. Since in principle more than one hyperfine level of the molecular ground state can be
populated in the STIRAP transfer, a natural question is how to quantitatively
characterize the purity of the hyperfine levels of the ground state. It seems that this question
cannot be answered by just observing the Feshbach molecules, since the
ground-state molecule cannot be directly detected. Another relevant question is, if the STIRAP does not prepare the molecule in a single
hyperfine level of the ground state, whether the molecule is in a coherent
superposition or in an incoherent mixture of different hyperfine states? To answer this question is
particularly important to the fermionic molecules, as the fermionic molecules in a superposition state are still identical fermions, and thus
\textit{s}-wave collisions among them are suppressed due to the Pauli principle \cite{Gupta2003,Zwierlein2003}. Incoherent mixtures, in
contrast, suffer from \textit{s}-wave collisions. Therefore whether the created state is a coherent superposition state or an incoherent mixture is important to interpret the lifetime of the ground state molecules \cite{Park2015}.

In this Letter, we report on the first observation of the interference between resonant and detuned STIRAP in the adiabatic
creation of $^{23}$Na$^{40}$K ground-state molecules. The interference manifests itself as the oscillation of the number of Feshbach molecules after a round-trip
STIRAP versus the hold time between the forward and the reverse
STIRAP. Surprisingly, the oscillation has a high
visibility of about $90\%$ even if the single-photon detuning of the detuned STIRAP is about one order of magnitude larger than the linewidth of the excited state and the Rabi frequencies of the STIRAP lasers. At the same time, the STIRAP spectrum shows two
peaks, and the spectrum changes with the hold time. The interference phenomenon indicates that if more than one hyperfine level of the ground state is populated, the STIRAP creates a superposition state, but not an incoherent mixure. Moreover, the interference provides a tool to
quantitatively characterize the purity of the hyperfine levels of the ground-state molecule.

Our experiment starts with $^{23}$Na$^{40}$K Feshbach molecules which are associated
from an ultracold atomic mixture in a crossed-beam optical dipole trap at a magnetic field of
$89.77(1)$ G, which is close to the atomic \textit{s}-wave Feshbach resonance
between $|f,m_{f}\rangle_{\mathrm{{Na}}}%
=|1,1\rangle$ and $|f,m_{f}\rangle_{\mathrm{{K}}}=|9/2,-7/2\rangle$ at $90.33$
G \cite{Park2012}. The binding energy of the Feshbach molecule is about $122$ kHz, and about
$2.3\times10^{4}$ Feshbach molecules can be formed. To transfer the molecules from the Feshbach
state $|F\rangle$ to the rovibrational ground state, we employ a mixture
of the electronic excited states $B^{1}\Pi$ $|v=12,J=1\rangle$ and $c^{3}\Sigma$ $|v=35,J=1\rangle$.
The hyperfine structure of the electronic excited state has
been extensively investigated in Refs. \cite{Ishikawa1992,Park2015b}.

\begin{figure}[ptb]
\centering
\includegraphics[width=8cm]{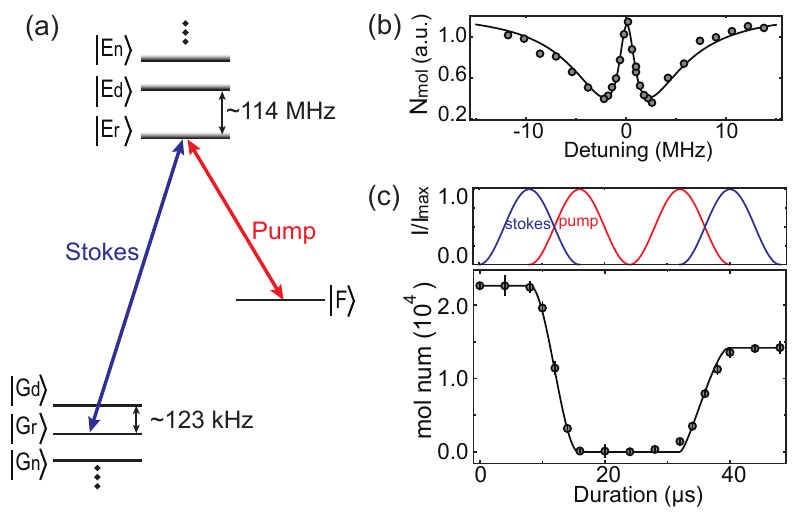}\caption{(a) Illustration of stimulated Raman adiabatic passage (STIRAP) in the adiabatic creation of the $^{23}$Na$^{40}$K molecule. The $|E_r\rangle$, $|E_d\rangle$ and $|E_n\rangle$ states are the hyperfine levels of the electronic excited state, and  the $|G_r\rangle$, $|G_d\rangle$ and $|G_n\rangle$ states are the hyperfine levels of the rovibrational ground state. The $|E_r\rangle$ state is well-separated from the other hyperfine excited states. The pump light resonantly couples the Feshbach state $|F\rangle$ and the intermediate electronic state $|E_r\rangle$, and the Stokes light resonantly couples the ground state $|G_r\rangle$ and $|E_r\rangle$. (b) Two-photon dark-state spectroscopy. (c) The pulse sequence of the round-trip STIRAP. The number of Feshbach molecules during the STIRAP process is measured by truncating the laser pulses.  Error bars represent $\pm1$ s.d. }
\label{fig1}%
\end{figure}

We choose the hyperfine level $|E_{r}\rangle$ of the electronic excited state
as the intermediate state, which may be approximately
described by the quantum numbers $F_{1}\approx1/2,m_{F_{1}}\approx
-1/2,m_{I_{K}}\approx-2$ and a total angular momentum projection $m_{F}%
=-5/2$. As shown in Fig. 1(a), the energy difference between this hyperfine state and the nearest
neighbouring hyperfine excited state is 114(10) MHz, which is much larger than the
linewidth $10(1)$ MHz of the excited state, and thus the $|E_{r}\rangle$
hyperfine state is well-separated from other hyperfine excited states. The pump laser coupling the $|F\rangle$
state to the $|E_{r}\rangle$ state, and the Stokes laser coupling $|E_{r}\rangle$ and the ground state are both $\pi-$polarized. Because the $|E_{r}\rangle$ state has
$m_{I_{\mathrm{{K}}}}\approx-2$ and $m_{F}%
=-5/2$, the hyperfine level of the ground state that is dominantly
populated by the Stokes light is given by $|G_{r}\rangle
=|v,N,m_{I_{\mathrm{{Na}}}},m_{I_{\mathrm{{K}}}}\rangle=|0,0,-1/2,-2\rangle$, where $v$ and $N$ are the vibrational and rotational quantum numbers.
In this case, we expect the $|E_r\rangle$, $|G_r\rangle$ and $|F\rangle$ states to form an almost ideal three-level system.

\begin{figure}[ptb]
\centering
\includegraphics[width=8cm]{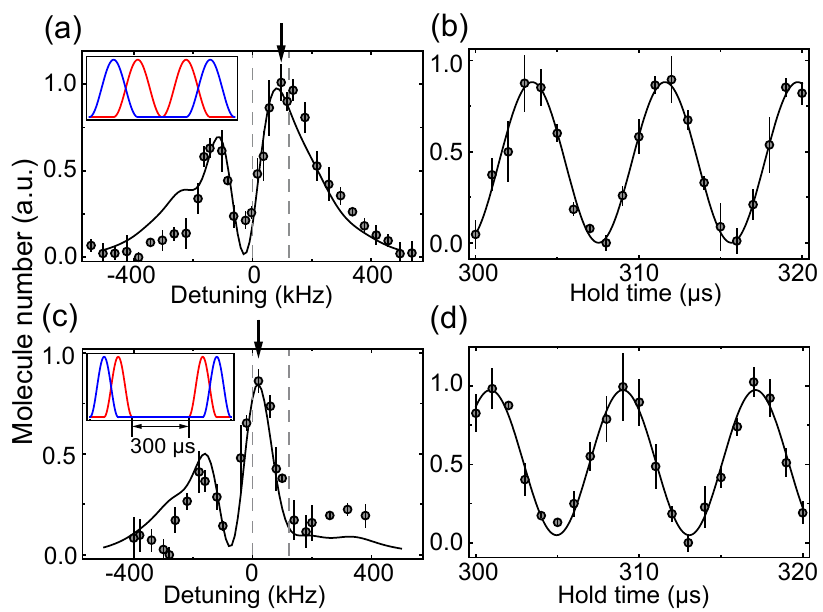}\caption{(a) The STIRAP spectrum is measured as a function of the frequency of the pump laser. The solid line is a theoretical fit to the shape of the data (see text). The two vertical dashed lines represent two-photon resonances between the Feshbach state $|F\rangle$ and the hyperfine levels $|G_r\rangle$ and $|G_d\rangle$ of the ground state, respectively. The pulse sequence is shown in the inset. (b) The number of Feshbach molecules after a round-trip STIRAP as a function of hold time between the forward and reverse STIRAP oscillates with a visibility of 90\% and a  frequency of about 123 kHz. The pump laser frequency used for this measurement is marked by the arrow in (a). (c) The STIRAP spectrum is measured for a hold time of 300 $\mu$s. The pulse sequence is shown in the inset. (d) The number of Feshbach molecules after a round-trip STIRAP as a function of hold time. The pump laser frequency used for this measurement is marked by the arrow in (c). Error bars represent $\pm1$ s.d.}
\label{fig2}%
\end{figure}

In the experiment, the pump laser (805 nm) and the Stokes
laser laser (567 nm) are locked to an ultralow
expansion (ULE) cavity. The peak Rabi frequency  of the pump light is
$\Omega_{P}=2\pi\times4.6(3)$ MHz, and the peak Rabi frequency  of the Stokes
light is $\Omega_{S}=2\pi\times9.3(4)$ MHz. They are determined from the two-photon
dark-state spectroscopy as shown in Fig. 1(b). We first perform a round-trip STIRAP using the pulse sequence in Fig. 1(c). The number of the remaining Feshbach molecules during the STIRAP
process is measured by truncating the pump and the Stokes laser pulses and detecting the
molecules in the $|F\rangle$ state. The results show that, starting from about
$2.3\times10^{4}$ Feshbach molecules, after a round-trip STIRAP about
$1.4\times10^{4}$ Feshbach molecules are recovered. This corresponds to a
one-way (round-trip) transfer efficiency of about $78\%$ ($61\%$), which indicates that about
$1.8\times10^{4}$ ground-state molecules are created.

So far, the experimental results are consistent with our understanding based on a simple three-level model. We then measure the STIRAP spectrum by scanning the frequency of the pump
laser. As shown in Fig. 2(a), we observe two well-separated peaks in the
spectrum, with a frequency difference of about $240$ kHz. The observation of
two peaks is unexpected, since the Stokes laser dominantly couples the
$|E_{r}\rangle$ state with the $|G_{r}\rangle$ state, and a numerical
calculation shows that the coupling strength between the $|E_{r}\rangle$ and
$|G_{r}\rangle$ states is about one order of magnitude larger than to other
hyperfine levels of the ground state.



We set the frequency of the pump laser to the higher peak and measure the
lifetime of the ground-state molecule by varying the hold time between forward and reverse STIRAP. We observe that the number of the
Feshbach molecules after a round-trip STIRAP oscillates as a function
of the hold time, with a high visibility of about $90\%$ and a frequency of
about $123$ kHz, as is shown in Fig. 2(b). This indicates the
efficiency of the reverse STIRAP oscillates with the hold time. Note that an oscillation of the Feshbach molecule number after a round-trip
transfer was also reported in Ref. \cite{Rvachov2017}. In that work, the oscillation frequency is
consistent with the axial trapping frequency, and the oscillation is
attributed to the axial sloshing. However, in our experiment, the oscillation
frequency is about three orders of magnitude larger than the trap frequencies.

To understand these phenomena, we measure the STIRAP spectrum using a different pulse sequence. As shown in Fig. 2(c), the
shape of the spectrum changes and the higher peak is shifted by about 60 kHz.
By setting the frequency of the pump laser to this higher peak, we measure the number of
Feshbach molecules as a function of the hold time, which still oscillates with the same frequency but with a different phase (see Fig. 2(d)).

These results strongly suggest that what we observe is a quantum interference
phenomenon. The oscillation frequency of about $123$ kHz is very close to the
energy difference between the $|G_{r}\rangle$ state and the adjacent hyperfine level of the
ground state $|G_{d}\rangle=|0,0,-3/2,-1\rangle$ at $B=89.77$ G \cite{Park2015,Aldegunde2017}. Therefore, these phenomena may be due to interference involving these
two hyperfine levels. However, the coupling strength between the $|E_{r}\rangle$ and $|G_{d}\rangle$ states is negligible, and thus the resonant STIRAP will not populate the $|G_d\rangle$ state.

\begin{figure}[ptb]
\centering
\includegraphics[width=8cm]{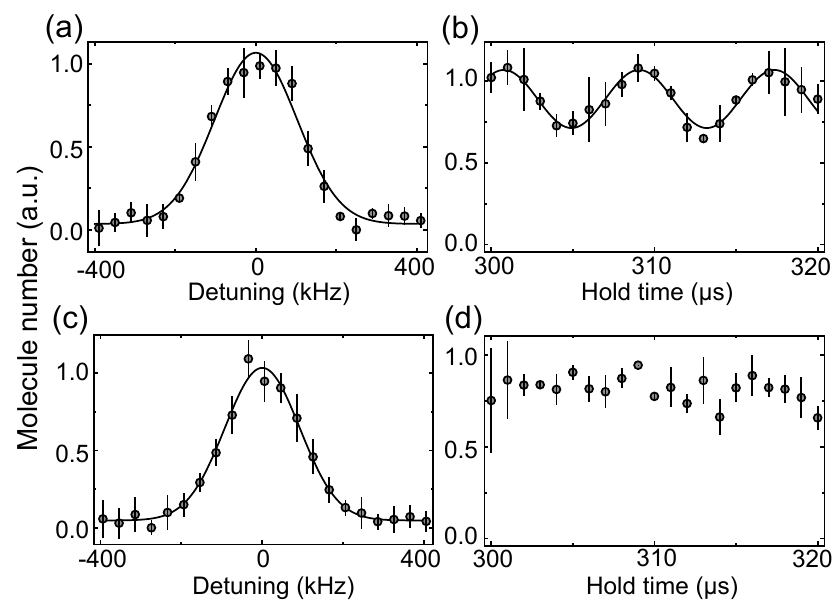}\caption{Increase of the purity by reducing the Rabi frequencies of the STIRAP lasers. (a) The STIRAP spectrum and (b) the measured oscillation with $\Omega_P, \Omega_S\approx2\pi\times4.6$ MHz. The estimated purity of the hyperfine ground state is about 90\%. (c) The STIRAP spectrum and (d) the measured oscillation with $\Omega_P, \Omega_S \approx2\pi\times3.3$ MHz. There is no obvious oscillation in (d). The estimated purity of the hyperfine level of the ground state is about 99\%.  Error bars represent $\pm1$ s.d.}
\label{fig3}%
\end{figure}

After carefully studying the molecule structure and the STIRAP process, we
find these phenomena can be explained by taking the
detuned STIRAP between the $|F\rangle$ and $|G_{d}\rangle$ states via the
neighbouring hyperfine excited state $|E_{d}\rangle$ into account, described by the quantum
numbers $F_{1}\approx3/2,m_{F_{1}}\approx-3/2,m_{I_{K}}\approx-1$. Because the
$|E_{d}\rangle$ state has $m_{I_{\mathrm{{K}}}}\approx-1$, the detuned STIRAP
dominantly transfers the molecule to the $|G_{d}\rangle$ state. Therefore, the interference between
the resonant and detuned STIRAP induces the coherence between $|G_r\rangle$ and $|G_d\rangle$. The
$|E_{d}\rangle$ level is higher than the $|E_{r}\rangle$ level by $114(10)$
MHz, and it is lower than the next higher excited state $|E_n\rangle$ by about $80$ MHz.
The single-photon detuning of the detuned STIRAP via the $|E_{d}\rangle$ is one order of magnitude larger than the Rabi
frequencies of the lasers and the linewidth of the excited state. For such a large
detuning, it is surprising that the detuned STIRAP cannot be
neglected if compared with the resonant STIRAP.

The observed oscillation may be understood as follows.
Starting from the $|F\rangle$ state, the resonant STIRAP transfers the molecules from the $|F\rangle$ to the $|G_{r}\rangle$ state,
while the detuned STIRAP transfers them from the
$|F\rangle$ to the $|G_{d}\rangle$ state. These two processes interfere
with each other, and consequently the forward STIRAP creates a superposition
state between $|G_{r}\rangle$ and $|G_{d}\rangle$, which may be described by
$|G\rangle_{in}=\cos\theta_{1}|G_{r}\rangle+e^{i\phi_{1}}\sin\theta_{1}|G_{d}\rangle$,
with $\theta_{1}$ and $\phi_{1}$ being unknown parameters. After the forward STIRAP, the
superposition state freely evolves according to $|G(t)\rangle_{in}=\cos
\theta_{1}|G_{r}\rangle+e^{i(2\pi\nu_{rd}t+\phi_{1})}\sin\theta_{1}|G_{d}\rangle$,
where $\nu_{rd}$ is the frequency difference between the $|G_{r}\rangle$
and the $|G_{d}\rangle$ state. Assuming that the reverse STIRAP converts the
$|G\rangle_{out}=\cos\theta_{2}|G_{r}\rangle+e^{i\phi_{2}}\sin\theta_{2}|G_{d}\rangle$ state back to the $|F\rangle$ state, we expect that the
conversion efficiency for a hold time of $t$ is proportional to $|\langle
G_{in}(t)|G_{out}\rangle|^{2}=\cos^{2}\theta_{1}\cos^{2}\theta_{2}+\sin^{2}\theta_{1}\sin^{2}\theta_{2}+
\sin2\theta_{1}\sin2\theta_{2}\cos(2\pi\nu_{rd}t+\phi_{1}-\phi_{2})/2$, which oscillates with a frequency
of $\nu_{rd}$.



To quantitatively explain the experimental results, we numerically calculate
the STIRAP spectrum \cite{supp}. In our calculation, we include three neighbouring hyperfine excited states $|E_i\rangle$ with $i=r,d,n$ and three
neighbouring hyperfine ground states $|G_i\rangle$ with $i=r,d,n$, which can be coupled by the Stokes and the pump light.
The relative coupling strengths between the different states are calculated by using the method introduced in Ref. \cite{Park2015b,seesselberg2018,supp}. The shape of the theoretical STIRAP spectrum is fitted to the experimental data, as shown in Fig. 2(a) and 2(c). The two peak
structures of the STIRAP spectrum are reproduced by the theoretical
calculations. The valley between the two peaks is caused by destructive
interference, and the peaks are shifted from two-photon resonances due to the interference. The numerical calculations show that
the forward STIRAP mainly creates a superposition between the $|G_r\rangle$ and $|G_d\rangle$ states, and that the efficiency of the round-trip STIRAP oscillates with a frequency equal to $\nu_{rd}$.

The observed interference demonstrates that in the STIRAP transfer, if more
than one hyperfine level of the ground state can be populated, the forward STIRAP
creates a superposition of the different hyperfine states. It is still a fully
polarized state, but not an incoherent mixture. For the
fermionic molecules, the \textit{s}-wave collision between the ground-state
molecules in a superposition state is still suppressed due to the Pauli principle \cite{Gupta2003,Zwierlein2003}.

\begin{figure*}[ptb]
\centering
\includegraphics[width=16cm]{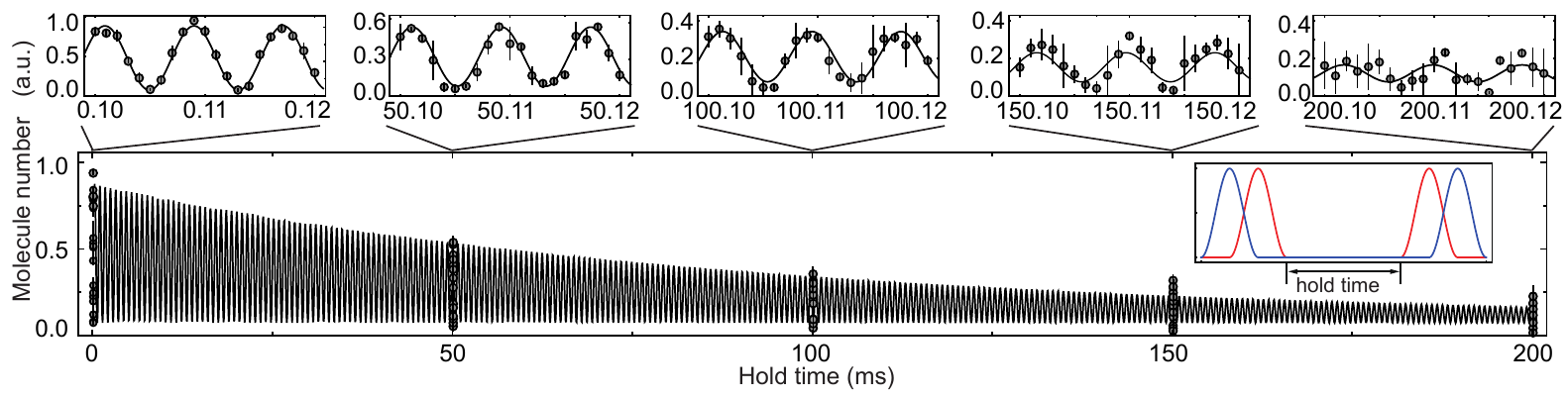}\caption{The coherence between the $|G_{r}\rangle$ and the $|G_{d}\rangle$ state is measured for a long hold time. We fit the data points with a single oscillation function, which gives a frequency of 123.080(1) kHz. The oscillation can be clearly observed for 150 ms, which corresponds to more than 10$^4$ oscillations. The pulse sequence is shown in the inset.  Error bars represent $\pm1$ s.d.}
\label{fig4}%
\end{figure*}

The observed oscillation provides a method to quantitatively
characterize the purity of the hyperfine levels of the created state. Assuming $\theta_{1}=\theta_{2}=\theta$, the visibility of the oscillation may be expressed as
$vis=(1-\cos4\theta)/(3+\cos4\theta)$. The visibility in our experiment
is about $0.90$, and a simple calculation yields $|G\rangle_{in}%
=\sqrt{0.61}|G_{r}\rangle+e^{i\phi_1}\sqrt{0.39}|G_{d}\rangle$. Therefore we
determine the purity of the ground-state molecule in the $|G_{r}\rangle$
state to be about $0.61$.

The purity can be increased by reducing the Rabi frequencies of the lasers. The STIRAP spectrum and the efficiency versus the hold time for different Rabi frequencies are plotted in Fig. 3. By reducing the Rabi frequencies to $\Omega_{P}, \Omega_{S}\approx$
$2\pi\times3.3$ MHz, no obvious
oscillation can be observed, as shown in Fig. 3(d). If we still fit the data points with an oscillation
function, we obtain a visibility of $0.03$, and thus the purity in
$|G_{r}\rangle$ is $0.99$, which is very high.

The created superposition state created is between the $|G_{r}\rangle$
and $|G_{d}\rangle$ states.
These two states have only nuclear spins,
and their frequency difference is insensitive to the magnetic field. We
measure the oscillation for a long hold time. As shown in Fig. 4, the oscillation can be
clearly observed after $150$ ms which corresponds to more than $10^{4}$ oscillations. By fitting the data points with a single oscillation
function, we obtain the frequency of $123.080(1)$ kHz. From this frequency, we determine the scalar coefficient $c_{4}=416(35)$ Hz in the molecule Hamiltonian, which is consistent with the result in Ref. \cite{Park2017}. The superposition between different hyperfine ground states may be employed to implement a STIRAP interferometer. The superposition state created in our work is due to the interference between resonant and detuned STIRAP. The superposition state may also be created by employing  polarizations of lasers. For example, a superposition state may be easily created in STIRAP by employing the $(\sigma^{+}+\sigma^{-})/\sqrt{2}$ polarized pump light to couple the Feshbach state to two hyperfine excited states, and by employing the $\pi$ polarized Stokes light to couple each hyperfine excited state to one hyperfine ground state.
Besides, a similar interferometer could be created by stopping STIRAP ¡°half-way¡±~\cite{Winkler2005,Semczuk2014}.

In summary, we have observed interference between resonant and detuned STIRAP in the adiabatic creation of $^{23}$Na$^{40}$K ground-state molecules.
Such an interference phenomenon adds an unexpected new basic aspect to STIRAP, which has been overlooked in the past. The interference clearly demonstrates that if more than one hyperfine state is populated, the state created by STIRAP is a superposition state, and thus the \emph{s}-wave collisions  between the fermionic molecules are still suppressed due to the Pauli principle. The superposition state can be changed to an incoherent mixture by introducing  decoherence mechanisms. The mixture might be useful in evaporative cooling, as demonstrated the $^{6}$Li atomic gases~\cite{Gupta2003,Ketterle2008}. The observed oscillation provides a method to quantitatively measure the purity of the ground-state molecules. The STIRAP interferometer may also be employed to detect collisions between molecules.

\begin{acknowledgments}
This work was supported by the National  Key R\&D Program of China (under Grant No. 2018YFA0306502), the National Natural Science Foundation of China (under Grant No. 11521063),  the Chinese Academy of Sciences, the Anhui Initiative in Quantum Information Technologies, and Shanghai Sailing Program (under Grant No. 17YF1420900). 
\end{acknowledgments}


\section{supplementary materials}

\section{A. STIRAP}

To quantitatively understand the experimental results, we use a seven-level model to numerically simulate the STIRAP process. In the supplementary materials, we denote the three hyperfine levels of the excited states $|E_r\rangle$, $|E_d\rangle$ and $|E_n\rangle$ by $|e_1\rangle$, $|e_2\rangle$ and $|e_3\rangle$, respectively,  and we denote the three hyperfine levels of the ground state $|G_r\rangle$, $|G_d\rangle$ and $|G_n\rangle$ by $|g_1\rangle$, $|g_2\rangle$ and $|g_3\rangle$ respectively. The Feshbach state $|F\rangle$ is denoted by $|f\rangle$. The pump light couples $|e_{1,2,3}\rangle$ to $|f\rangle$, and the Stokes light couples $|e_{1,2,3}\rangle$ to $|g_{1,2,3}\rangle$. In the rotating frame, the Hamiltonian governing the STIRAP process is given by
\begin{widetext}
\begin{equation}
\renewcommand\theequation{S\arabic{equation}}
H(t)=\left(
  \begin{array}{ccccccc}
    \delta_f & \frac{\Omega_{e_1 f}(t)}{2} & \frac{\Omega_{e_2 f}(t)}{2} & \frac{\Omega_{e_3 f}(t)}{2} & 0 & 0 & 0 \\
    \frac{\Omega_{e_1 f}(t)}{2} & \Delta_{e_1}-i \frac{\Gamma}{2} & 0 & 0 & \frac{\Omega_{e_1 g_1}(t)}{2} &  \frac{\Omega_{e_1 g_2}(t)}{2} &  \frac{\Omega_{e_1 g_3}(t)}{2} \\
    \frac{\Omega_{e_2 f}(t)}{2} & 0 & \Delta_{e_2}-i \frac{\Gamma}{2} & 0 &  \frac{\Omega_{e_2 g_1}(t)}{2} &  \frac{\Omega_{e_2 g_2}(t)}{2} &  \frac{\Omega_{e_2 g_3}(t)}{2} \\
    \frac{\Omega_{e_3 f}(t)}{2} & 0 & 0 & \Delta_{e_3}-i \frac{\Gamma}{2} & \frac{\Omega_{e_3 g_1}(t)}{2} & \frac{\Omega_{e_3 g_2}(t)}{2} & \frac{\Omega_{e_3 g_3}(t)}{2} \\
    0 & \frac{\Omega_{e_1 g_1}(t)}{2} & \frac{\Omega_{e_2 g_1}(t)}{2} & \frac{\Omega_{e_3 g_1}(t)}{2} & \delta_{g_1} & 0 & 0 \\
    0 &\frac{\Omega_{e_1 g_2}(t)}{2} & \frac{\Omega_{e_2 g_2}(t)}{2} & \frac{\Omega_{e_3 g_2}(t)}{2} & 0 & \delta_{g_2} & 0 \\
    0 & \frac{\Omega_{e_1 g_3}(t)}{2} & \frac{\Omega_{e_2 g_3}(t)}{2} & \frac{\Omega_{e_3 g_3}(t)}{2} & 0 & 0 & \delta_{g_3} \\
  \end{array}
\right)
\end{equation}
\end{widetext}
In the case that the pump light resonantly couples $|e_1\rangle$ to $|f\rangle$, and the Stokes light resonantly couples $|e_1\rangle$ to $|g_1\rangle$, we have $\Delta_{e_1}=\delta_{g_1}=\delta_{f}=0$.
The linewidth of the excited state $\Gamma=2\pi\times10$ MHz, the detuning of the other two excited states $\Delta_{e_2}=11.4 \Gamma$, $\Delta_{e_3}=19.6\Gamma$, and the peak Rabi frequencies $\Omega_{e_1 f}=\Omega_P=0.46\Gamma$, $\Omega_{e_1 g_1}=\Omega_S=0.93\Gamma$ can be determined in the experiment. The other parameters are determined by theoretical calculations. The detuning of the other two hyperfine levels of the ground state are $\delta_{g_2}=0.012308\Gamma$ and $\delta_{g_3}=-0.012224\Gamma$. The values of the relative coupling strengths are listed in Table S1. The details of the calculations are given in Sec. B, C, and D of the supplementary materials.
\begin{table}[pth]
\renewcommand\thetable{S1}
\begin{tabular}{|c|c|}\hline
  $\Omega_{e_2 f}/\Omega_{e_1 f}$ & 0.712 \\ \hline
  $\Omega_{e_3 f}/\Omega_{e_1 f}$ & 0.416 \\ \hline
  $\Omega_{e_1 g_2}/\Omega_{e_1 g_1}$& -0.144    \\ \hline
  $\Omega_{e_1 g_3}/\Omega_{e_1 g_1}$& -0.009    \\ \hline
  $\Omega_{e_2 g_1}/\Omega_{e_1 g_1}$& 0.023     \\ \hline
  $\Omega_{e_2 g_2}/\Omega_{e_1 g_1}$& -1.16     \\ \hline
  $\Omega_{e_2 g_3}/\Omega_{e_1 g_1}$& 0.0455    \\ \hline
  $\Omega_{e_3 g_1}/\Omega_{e_1 g_1}$& 0.514     \\ \hline
  $\Omega_{e_3 g_2}/\Omega_{e_1 g_1}$& 0.199     \\ \hline
  $\Omega_{e_3 g_3}/\Omega_{e_1 g_1}$& 0.297    \\ \hline
\end{tabular}
\caption{The calculated relative coupling strengths. }
\label{Table}
\end{table}

In our experiment, the time dependent intensity of the pump and Stokes light is approximately given by
\begin{equation}
\renewcommand\theequation{S\arabic{equation}}
\frac{I(t)}{I_{max}}=\frac{1-\cos(\pi (t-t_{i})/t_d)}{2}
\end{equation}
for $t_{i}<t<t_i+2 t_d$ with $t_d=8$ $\mu$s, which is obtained by controlling the rf power driving the acoustic-optical modulator.

Given these parameters, we can numerically simulate the STIRAP process by solving the time dependent Schr\"{o}dinger equation
\begin{equation}
\renewcommand\theequation{S\arabic{equation}}
i|\dot{\psi}(t)\rangle=H(t)|\psi(t)\rangle,
\end{equation}
where the wave function is expressed as
\begin{equation}
\renewcommand\theequation{S\arabic{equation}}
|\psi(t)\rangle=c_f(t)|f\rangle+\sum_{i}c_{g_{i}}(t)|g_{i}\rangle+\sum_{j}c_{e_{j}}(t)|e_{j}\rangle
\end{equation}
with the initial condition $c_f(0)=1$ and $c_{g_{1,2,3}}(0)=c_{e_{1,2,3}}(0)=0$.

The number of Feshbach molecules after a round-trip STIRAP is proportional to the probability in the Feshbach state $P_f(\infty)=|c_f(\infty)|^2$. In Fig. S1, We plot the calculated $P_f(\infty)$ as a function of hold time. The dashed line is the theoretical curve for the Rabi frequencies $\Omega_S=0.93\Gamma$ and $\Omega_P=0.46\Gamma$. The oscillation frequency is given by 123.08 kHz, which is equal to the frequency difference between $|g_1\rangle$ and $|g_2\rangle$. The calculated STIRAP efficiency is higher than the experimental result, which may be due to the remaining phase noises of the pump and Stokes lasers.

The STIRAP spectrum is given by plotting $P_f(\infty)$ as a function of $\delta_f$. The spectrum show two peak structures and is also a period function of the hold time, which is consistent with the experimental results. The shape of the calculated spectrum is fitted to the experimental data, as shown in the Fig. 2(a) and 2(c) of the main text.

\begin{figure}[ptb]
\centering
\renewcommand\thefigure{S1}
\includegraphics[width=8cm]{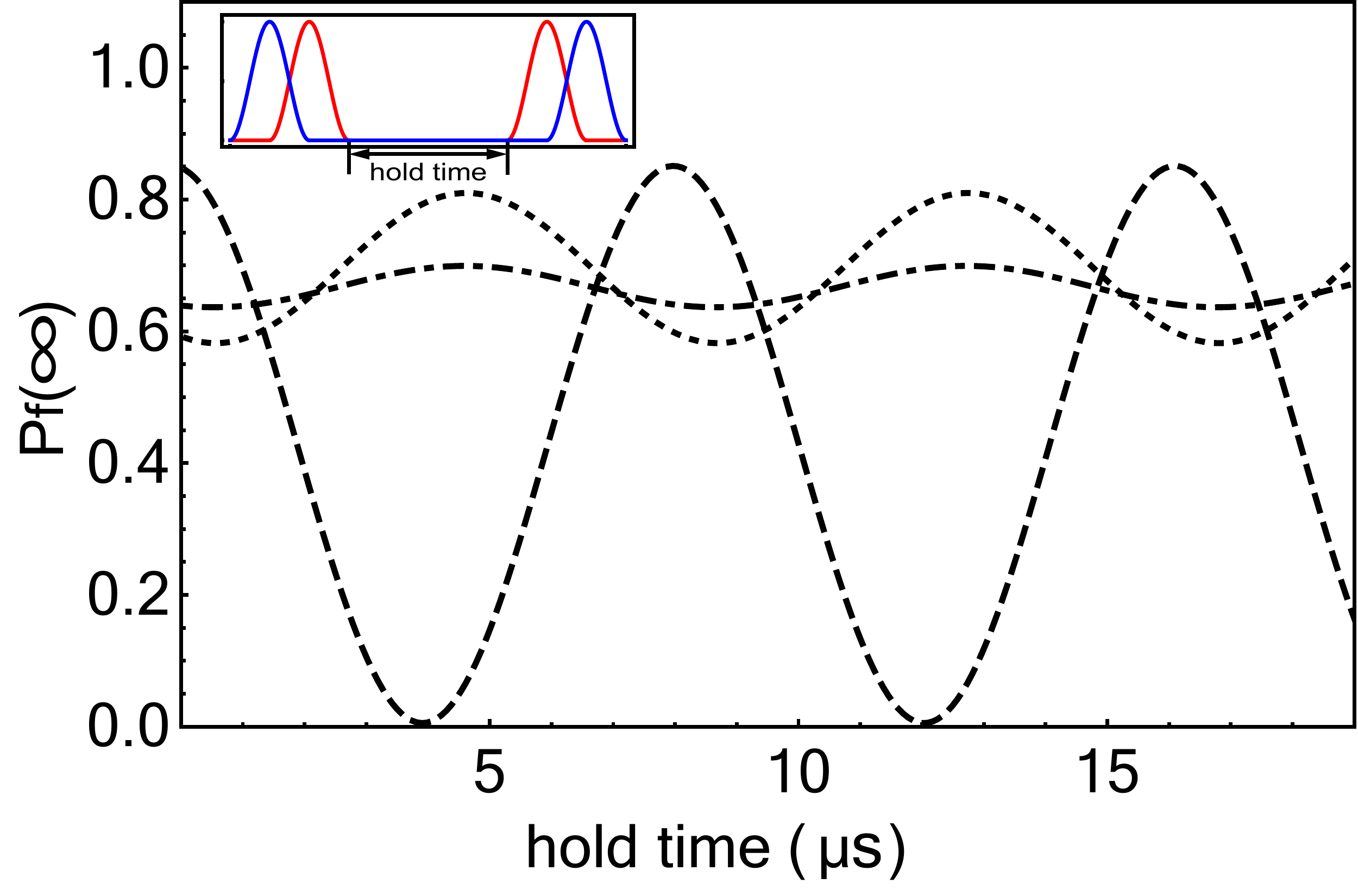}
\caption{The probability in the Feshbach state after a round-trip STIRAP $P_f(\infty)=|c_f(\infty)|^2$  as a function of the hold time  are calculated using the theoretical model. The parameters used in the calculations are $\Omega_S=0.93\Gamma, \Omega_P=0.46 \Gamma$ (dashed line), $\Omega_S=\Omega_P=0.46 \Gamma$ (dotted line), and $\Omega_S=\Omega_P=0.33 \Gamma$ (dotdashed line). The inset is the pulse sequence. } %
\label{figs1}%
\end{figure}

Given the theoretical model, we find that the effect of interference can be suppressed by reducing the Rabi frequencies of the pump and Stokes lasers. As shown in Fig. S1, by reducing the Rabi frequencies to $\Omega_S=\Omega_P=0.33\Gamma$, the interference is largely suppressed. In our experiment, when reducing the Rabi frequencies to  $\Omega_S, \Omega_P\approx0.33\Gamma$, no obvious oscillations can be observed, as shown in Fig. 3 of the main text.

In the following sections, we will give the details of the calculations to determine the energy differences $\delta_{g_{2}}$ and $\delta_{g_{3}}$ and the relative coupling strengths listed in Table S1.

\section{B. ground state}

The rovibrational ground state has good quantum numbers $N=S=J=M_J=0$. The quantum state of the hyperfine level of the ground state can be expressed as $|NSJM_Jm_{i_{\rm{Na}}}m_{i_{\rm{K}}}\rangle$, where the projections of nuclear spins $m_{i_{\rm{Na}}}$ and $m_{i_{\rm{K}}}$ are also approximately good quantum numbers at the large magnetic fields. The energy difference $\delta_{g_{2}}$ and $\delta_{g_{3}}$ of different hyperfine levels are calculated using the Hamiltonian \cite{Aldegunde2017}
\begin{equation}
\renewcommand\theequation{S\arabic{equation}}
H_g=g_{i_{\rm{Na}}}\mu_{B} i_{z_{\rm{Na}}}B_z+g_{i_{\rm{K}}}\mu_{B} i_{z_{\rm{K}}}B_z-c_4\mathbf{i}_{\rm{Na}}\cdot\mathbf{i}_{\rm{K}}
\end{equation}
where $g_{i_{\rm{Na}}}=-0.00080461$ and $g_{i_{\rm{K}}}=0.00017649$  are the nuclear g-factors and $B_z$ is the bias magnetic field. We have used the measured value $c_4=416$ Hz in the calculation.

\section{C. Feshbach state}
To calculate the relative coupling strengths between the different hyperfine levels of the excited states and the Feshbach state, we first calculate the wave function of the Feshbach state using the coupled-channel method.
The Hamiltonian describing the $s$-wave bound state is
\begin{equation}
\renewcommand\theequation{S\arabic{equation}} H=T+\sum_{S=0,1}V_{S}%
(r)P_{S}+H_{hf}+H_{z}.
\end{equation}
The first term is the kinetic energy $T=-\frac{\hbar^{2}}{2\mu}\frac{d^{2}}{ dr^{2}}$
with $\mu$ the reduced mass. The second term describes the spin-exchanging
interaction, where $P_{0}=1/4-\mathbf{s}_{\rm{Na}}\cdot\mathbf{s}_{\rm{K}}$ and
$P_{1}=3/4+\mathbf{s}_{\rm{Na}}\cdot\mathbf{s}_{\rm{K}}$ are the singlet and
triplet projection operator respectively with $\mathbf{s}$ the electron spin.
$V_{0}(r)$ and
$V_{1}(r)$ denotes the Born-Oppenheimer singlet potential $X^{1}\Sigma$ and
triplet potential $a^{3}\Sigma$. The Born-Oppenheimer potentials can be
expressed as power expansions of $r$, whose exact form can be found in Ref.
\cite{Temelkov2015,Zhu2017}. $H_{hf}$ is the hyperfine interaction term, described by%
\begin{equation}
\renewcommand\theequation{S\arabic{equation}} H_{hf}=a_{\rm{Na}}%
\mathbf{s}_{\rm{Na}}\cdot\mathbf{i}_{\rm{Na}}+a_{\rm{K}}\mathbf{s}_{\rm{K}}%
\cdot\mathbf{i}_{\rm{K}},
\end{equation}
where $a_{\rm{Na,K}}$ is the hyperfine constant and $\mathbf{i}$ is the nuclear spin.
The last term is the Zeeman term
\begin{equation}
\renewcommand\theequation{S\arabic{equation}} H_{z}=[(g_{s}s_{z\rm{Na}
}-g_{i_{\rm{Na}}}i_{z\rm{Na}})+(g_{s}s_{z\rm{K}}-g_{i_{\rm{K}}}i_{z\rm{K}})]\mu
_{B}B_z,
\end{equation}
with $g_{s}$ the electron g-factor, $g_{i}$ the nuclear g-factor and $B_{z}$
the bias magnetic field.

The internal state may be expressed in terms of the spin basis $|\sigma
\rangle=|S, M_S, m_{i_{\rm{Na}}},m_{i_{\rm{K}}}\rangle.$
The Hamiltonian couples all the internal states with the same $M_{F}%
=M_S+m_{i_{\rm{Na}}}+m_{i_{\rm{K}}}$. For a given $M_{F}$
and $B_{z}$, we first diagonalize the $H_{hf}+H_{z}$ to obtain the internal
eigenstate $|\chi_{i}\rangle$ and the threshold energy $E_{i}^{th}$ of each
channel. Expanding the wave function in terms of the new bases $|\psi_f
\rangle=\sum_{i}\psi_{i}(r)|\chi_{i}\rangle$, we obtain the coupled channel
Schr\"{o}dinger equation
\begin{equation}
\renewcommand\theequation{S\arabic{equation}} \sum_{j}[T\delta_{ij}%
+\sum_{S=0,1}V_{S}(r)\langle\chi_{i}|P_{S}|\chi_{j}\rangle]\psi_{j}%
(r)=(E-E_{i}^{th})\psi_{i}(r)
\end{equation}
The binding energy and the wave function of the Feshbach state are calculated by means of the renormalized Numerov method \cite{Johnson1978}.

The closed-channel component of the Feshbach state is dominated by the highest bound state in the triplet potential. This bound state has good quantum numbers $N=0, S=1$ and $J=1$. The closed-channel component of the wave function of the Feshbach state can be expressed as
\begin{equation}
\renewcommand\theequation{S\arabic{equation}}
|\psi_f\rangle\propto\sum _{M_J m_{i_{\rm{Na}}} m_{i_{\rm{K}}}} c^f_{NSJ M_J m_{i_{\rm{Na}}} m_{i_{\rm{K}}} } |NSJ,M_J m_{i_{\rm{Na}}} m_{i_{\rm{K}}}\rangle.
\end{equation}
For the Feshbach state between $^{23}$Na $|1,1\rangle$ and
$^{40}$K $|9/2,-7/2\rangle$ at 89.77 G with a binding energy of 122 kHz, the calculated coefficients in the uncoupled basis are


\begin{table}[pth]
\begin{tabular}{|c|c|c|c|}\hline
   $M_J$ &$ m_{i_{\rm{Na}}} $&$ m_{i_{\rm{K}}}$&$  c^f_{NSJ,M_J, m_{i_{\rm{Na}}}, m_{i_{\rm{K}}} }$ \\ \hline
    -1 & -3/2    & 0             & -0.149 \\ \hline
    -1 & -1/2    & -1             & 0.447 \\ \hline
    -1 &  1/2    & -2             & -0.291 \\ \hline
    -1 &  3/2    & -3             & -0.154 \\ \hline
    0  & -3/2    & -1             & -0.194 \\ \hline
    0  & -1/2    & -2             & -0.236 \\ \hline
    0  &  1/2    & -3             & 0.156 \\ \hline
     0 &  3/2    & -4             & 0.270 \\ \hline
     1 & -3/2    & -2             & 0.631 \\ \hline
     1 & -1/2    & -3             & -0.266 \\ \hline
     1 &  1/2    & -4             & -0.103 \\ \hline
\end{tabular}
\label{Table}
\end{table}

\section{D. excited state and coupling strength}

The hyperfine structures of the excited states have been discussed in details in Ref~\cite{Park2015b,Ishikawa1992}. The excited state is a combination of a mixture of the electronic excited states $B^{1}\Pi$ $|v=12,J=1\rangle$ and $c^{3}\Sigma$ $|v=35,J=1\rangle$, and thus can be expressed as $|\psi_e\rangle=|\psi _{B^{1}\Pi}\rangle+|\psi _{c^{3}\Sigma}\rangle$.

The triplet component $|\psi _{c^{3}\Sigma}\rangle$ accounts for the coupling between the Feshbach state and excited state. The good quantum numbers are $ N'=1, S'=1$ and $ J'=1$, and thus $|\psi _{c^{3}\Sigma}\rangle$  can be expanded in the uncoupled basis $|N'S'J' M'_J m'_{i_{\rm{Na}}}m'_{i_{\rm{K}}}\rangle$.  The hyperfine and Zeeman interactions for $c^{3}\Sigma$ state are described by
\begin{equation}
\renewcommand\theequation{S\arabic{equation}}
H^c=a^{e}_{\rm{Na}}\mathbf{i}_{\rm{Na}}\cdot\mathbf{S}+a^{e}_{\rm{K}}\mathbf{i}_{\rm{K}}\cdot\mathbf{S}+g_s\mu_{B}B_z
\end{equation}

The singlet component $|\psi _{B^{1}\Pi}\rangle$ is responsible for the coupling between the excited state and the ground state. The good quantum numbers are $ \Lambda'=1, \Sigma'=0, \Omega'=1$, and $J'=1 $, and thus $|\psi _{B^{1}\Pi}\rangle$ can be expanded in the uncoupled basis $|J' \Omega' M'_J m'_{i_{\rm{Na}}}m'_{i_{\rm{K}}}\rangle$. The $B^{1}\Pi$ state has negligible hyperfine interaction and only the Zeeman interaction is important, which is described by
$\langle J' M'_J|H^B|J'M'_J\rangle=\mu_B M'_J B_z/(J'(J'+1))$. The $B^{1}\Pi$ and $c^{3}\Sigma$ states are mixed due to the spin-orbit coupling. The energy and the wave function  of the hyperfine levels of the excited states are  obtained by diagonalizing the coupled Hamiltonian.

The triplet component of the wave function of the hyperfine levels of the excited state can be expressed as
\begin{equation}
\renewcommand\theequation{S\arabic{equation}}
|\psi _{c^{3}\Sigma}\rangle \propto \sum_{M'_J m'_{i_{\rm{Na}}}m'_{i_{\rm{K}}}} c^e_{N'S'J' M'_J m'_{i_{\rm{Na}}}m'_{i_{\rm{K}}}}|N'S'J' M'_J m'_{i_{\rm{Na}}}m'_{i_{\rm{K}}}\rangle.
\end{equation}
Therefore the coupling strength between the excited state and the Feshbach state for the $\pi$-polarized pump light can be expressed as \cite{Brown2003}
\begin{widetext}
\begin{align}
\Omega_{ef}&\propto \langle \psi _{c^{3}\Sigma} |d_0 | \psi_f\rangle  \notag\\
 & \propto \sum_{M'_J m'_{i_{\rm{Na}}}m'_{i_{\rm{K}}}M_J m_{i_{\rm{Na}}}m_{i_{\rm{K}}}} c^{e}_{N'S'J' M'_J m'_{i_{\rm{Na}}}m'_{i_{\rm{K}}}} c^{f}_{NSJ M_J m_{i_{\rm{Na}}} m_{i_{\rm{K}}}}  \langle N'S'J' M'_J |d_0|NSJ M_J \rangle \delta_{m'_{i_{\rm{Na}}}m_{i_{\rm{Na}}}}\delta_{m'_{i_{\rm{K}}}m_{i_{\rm{K}}}} \notag\\
&\propto \sum_{M_J m_{i_{\rm{Na}}}m_{i_{\rm{K}}}} c^e_{N'S'J' M_J m_{i_{\rm{Na}}}m_{i_{\rm{K}}}} c^f_{NSJ M_J m_{i_{\rm{Na}}} m_{i_{\rm{K}}}}(-1)^{J'-M_J}
\left(                                                                           \begin{array}{ccc}
J' & 1 & J \tag{S13}\\                                                                             M_J & 0 & -M_J \\                                                                           \end{array}                                                                         \right).
\end{align}
\end{widetext}

The  singlet component of the wave function can be expressed as
\begin{equation}
|\psi _{B^{1}\Pi}\rangle\propto \sum _{M'_J m'_{i_{\rm{Na}}}m'_{i_{\rm{K}}}} c^e_{J'\Omega'  M'_J m'_{i_{\rm{Na}}}m'_{i_{\rm{K}}}}|J' \Omega' M'_J m'_{i_{\rm{Na}}}m'_{i_{\rm{K}}}\rangle. \tag{S14}
\end{equation}
Therefore the coupling strength between the excited state and the ground state $|\psi_g\rangle=|NSJM_Jm_{i_{\rm{Na}}}m_{i_{\rm{K}}}\rangle$ for the $\pi$-polarized Stokes light is described as \cite{Brown2003}
\begin{widetext}
\begin{align}
\Omega_{eg}&\propto \langle \psi _{B^{1}\Pi} |d_0 | \psi_g\rangle  \notag\\
 & \propto \sum_{M'_J m'_{i_{\rm{Na}}}m'_{i_{\rm{K}}}} c^{e}_{J'\Omega' M'_J m'_{i_{\rm{Na}}}m'_{i_{\rm{K}}}}   \langle J'\Omega' M'_J |d_0|NSJ M_J \rangle \delta_{m'_{i_{\rm{Na}}}m_{i_{\rm{Na}}}}\delta_{m'_{i_{\rm{K}}}m_{i_{\rm{K}}}} \notag\\
&\propto  c^e_{J'\Omega' M_J m_{i_{\rm{Na}}}m_{i_{\rm{K}}}} (-1)^{J'-M_J}
\left(                                                                           \begin{array}{ccc}
J' & 1 & J \tag{S15} \\                                                                             M_J & 0 & -M_J \\                                                                           \end{array}                                                                         \right)
\end{align}
\end{widetext}
The relative coupling strengths listed in Table S1 are calculated using Eqs. (S13) and (S15).

\end{document}